\documentclass[12pt,jpa]{iopart}
\usepackage{iopams}  
\usepackage{enumitem}
   
\usepackage{epsfig}
\usepackage{bm}
\usepackage{multirow}
\usepackage[colorlinks]{hyperref}
\usepackage{relsize}
\usepackage{xcolor}
\usepackage{cite} 

\hypersetup{colorlinks=true,
	linkcolor=red,
	anchorcolor=black,
	citecolor=blue,
	urlcolor=black
}

\usepackage{geometry}
\providecommand{\be}{\begin{equation}}
\providecommand{\ee}{\end{equation}}
\providecommand{\bea}{\begin{eqnarray}}
\providecommand{\eea}{\end{eqnarray}}
\providecommand{\beas}{\begin{eqnarray*}}
\providecommand{\eeas}{\end{eqnarray*}}
\providecommand{\bal}{\begin{aligned}}
\providecommand{\eal}{\end{aligned}}

\providecommand{\bsi}{{\bm \sigma}}

\begin{document}

\title[]{Reply to comment on `Real-space renormalization-group methods for hierarchical spin glasses'}
\author{Michele Castellana}

\address{Laboratoire Physico-Chimie Curie, Institut Curie, PSL Research University, CNRS UMR 168, Paris, France\\
Sorbonne Universit\'es, UPMC Univ. Paris 06, Paris, France}
\ead{michele.castellana@curie.fr}
\vspace{10pt}

\begin{abstract}
In their comment, Angelini et al. object to the conclusion of [\textit{J. Phys. A: Math. Theor.}, 52(44):445002, 2019], where we show that in [\textit{Phys. Rev. B}, 87(13):134201, 2013] the exponent $\nu$ has been obtained by applying a mathematical relation in a regime where this relation is not valid. 
We observe that the criticism above on the mathematical validity of such relation has not been addressed in the comment. Our criticism thus remains valid, and disproves the conclusions of the comment. This constitutes the main point of this reply. 

In addition, we provide a point-by-point response and discussion of Angelini et al.'s claims. First, Angelini et al. claim that the prediction $2^{1/\nu}=1$ of [\textit{J. Phys. A} 2019] is incorrect, because it results from the relation  $\lambda_{\rm max}=2^{1/\nu}$ between the largest eigenvalue of the linearized renormalization-group (RG) transformation and $\nu$, which cannot be applied to the ensemble renormalization group (ERG) method, given that for the ERG $\lambda_{\rm max} =1 $. However, the feature $\lambda_{\rm max} = 1$ is specific to the ERG transformation, and it does not give any grounds for questioning the validity of the general relation $\lambda_{\rm max}=2^{1/\nu}$ specifically for the ERG transformation. 
Second, Angelini et al. claim that $\nu$ should be extracted from an early RG regime (A), as opposed to the asymptotic regime (B) used to estimate $\nu$ in [\textit{J. Phys. A} 2019], and that (B) is dominated by finite-size effects. However, (A) is a small-wavelength, non-critical regime, which cannot characterize the critical exponent $\nu$ related to the divergence of the correlation length. Also, the fact that (B) involves finite-size effects is a feature specific to the ERG, and gives no rationale for extracting $\nu$ from (A). 

Finally, we refute the remaining claims made by Angelini et al. As a result of our analysis, we stand by our assertion that the ERG method yields a prediction given by $2^{1/\nu}=1$. 
\end{abstract}

\maketitle

\setlength{\parskip}{5pt plus 0pt minus 0pt}

In \cite{angelini2019comment}, Angelini et al. write that the estimate of $\nu$ made in \cite{angelini2013ensemble} relies on the relation 
\be\label{eq1}
\Delta_{t, 1}\equiv \frac{\beta_1 \sigma^{t\,1}_1 - \beta_2 \sigma^{t\, 2}_1}{\beta_1 - \beta_2} \propto 2^{t/\nu},
\ee
where  $\sigma^{t\,1}_1$ and $\sigma^{t\,2}_1$ are the standard deviations of the spin couplings at the first hierarchical level of the hierarchical Edwards-Anderson model (HEA), at the $t$-th ensemble renormalization group (ERG) step, and at inverse temperatures $\beta_1$ and $\beta_2$, respectively. 
However, in \cite{castellana2019real} we have shown that
\be\label{eq2}
\Delta_{t, 1} =  \sum_l (\lambda_l)^t \, v_{\rm R\, 1}^l ({\bm v}_{\rm L}^l \cdot  \bsi^0),
\ee
where ${\bm v}_{\rm L}$ and ${\bm v}_{\rm R}$ are the left and right eigenvectors of the matrix that linearizes the renormalization-group (RG) transformation at the critical fixed point, and $\bm \sigma^0 = (\sigma^0_1, \cdots, \sigma^0_n)$ the standard deviations of the spin couplings on hierarchical levels $1,\cdots,n$ at the beginning of the RG iteration.  If $t$ is large, i.e., large enough that the eigenvalue with the largest norm dominates the sum in Eq. (\ref{eq2}), then the relation (\ref{eq2})  reduces to
\be\label{eq4}
\Delta_{t, 1} =  [ v^1_{\rm R\, 1} ({\bm v}_{\rm L}^1 \cdot  \bsi^0)] \;  (\lambda_1)^t \hspace{1cm} (\textrm{large } t),
\ee
where $\lambda_1$ and ${\bm v}^1_{\rm L,R}$ are the (marginally) relevant eigenvalue and the corresponding left and right eigenvectors, respectively. As a result, for large $t$, Eq. (\ref{eq1}) is correct, with $\lambda_1 = 2^{1/\nu}$.
On the other hand,  Eq. (\ref{eq2})  demonstrates that Eq. (\ref{eq1})  does not hold if $t$ is small, i.e., small enough that multiple eigenvalues contribute to Eq. (\ref{eq2}), see ``[h]owever, we observe that the exponential dependence ... for $t \leq 3$, is incorrect'' in \cite{castellana2019real}.
The fact that the exponential form (\ref{eq1}) can hold for large $t$ only has been shown also in \cite{wilson1974the}---see ``[i]f $\lambda_1$ is the eigenvalue ... then for large $n$'' on page 110 and Eqs. (4.33), (4.37) and (4.38) therein.

Importantly, our criticism above on the mathematical grounds and validity of Eq. (\ref{eq1}) for small $t$, which was raised in \cite{castellana2019real}, has not been addressed in \cite{angelini2019comment}. As a result, this criticism remains valid, and refutes the conclusions of \cite{angelini2019comment}. This constitutes the main point of this reply. \vspace{1cm}\\

In addition, we comment on the use of the exponential form (\ref{eq4}) in \cite{angelini2013ensemble}. We observe that Eq. (\ref{eq4}) involves a proportionality factor $  v^1_{\rm R\, 1} ({\bm v}_{\rm L}^1 \cdot  \bsi^0)$ which is, in general, different from one, and which corresponds to the value in Fig. 5a of \cite{castellana2019real} which is reached by
$\sum_l (\lambda_l)^t \,v_{\rm R\, 1}^l ({\bm v}_{\rm L}^l \cdot  \bsi^0)$ for large $t$. 
On the other hand, the actual relation which has been used in \cite{angelini2013ensemble} to estimate $\nu$ is 
\be\label{eq3}
\Delta_{t, 1} =  2^{\frac{t}{\nu}},
\ee
see ``[t]o extract critical exponents ... in our case'' and ``[t]he procedure used is the same as that in the FM case'' \cite{angelini2013ensemble}. The relation (\ref{eq3}) differs from Eq. (\ref{eq4}) in the prefactor in the right-hand side which, in general, is different from one in Eq. (\ref{eq4}), and equals one in Eq. (\ref{eq3}). If Eq. (\ref{eq4}) were erroneously applied to the small-$t$ regime, where it is not valid,  then for $t=0$ its left-hand side would equal one, while its right-hand side would differ from one, thus resulting in a contradiction. To circumvent this inconsistency when improperly applying Eq. (\ref{eq3}) to the small-$t$ regime, its  prefactor $ v^1_{\rm R\, 1} ({\bm v}_{\rm L}^1 \cdot  \bsi^0)$ needs to be manually altered, and set to one. \\
Overall, the analysis above reflects the fact that, while Eq. (\ref{eq4}) has been formally derived from the RG equations \cite{castellana2019real}, Eq. (\ref{eq3}) lacks a mathematical justification. With reference to such a lack of mathematical grounds, in Section \ref{s4} we will show how the improper use of the exponential form (\ref{eq3}) for small $t$ in \cite{angelini2013ensemble} leads to an arbitrariness in the estimate of $\nu$. \\

In addition to the discussion above, in what follows we will present a point-by-point response and discussion of the remaining claims made in \cite{angelini2019comment}. 

\section{Finite-size effects}\label{s1}

In this Section we will address the criticisms raised in \cite{angelini2019comment} concerning the conclusions of \cite{castellana2019real} and finite-size effects in the ERG. 
We recall that the ERG method is composed of the following steps \cite{angelini2019comment}:
\begin{enumerate}[label=\arabic*)]
\item \label{p1} A HEA with $n$ hierarchical levels (a) and a HEA $n-1$ levels (b) are considered. Couplings at the $i$-th hierarchical level of (a) and (b) are normally distributed, with zero mean and standard deviation $\sigma_i$ and $\sigma'_i$, respectively.  
\item \label{p2} A set of observables for (a) is computed.
\item \label{p3} The standard deviations ${\bm \sigma}'=(\sigma'_1, \cdots, \sigma'_{n-1})$ of the couplings of (b) are determined by matching a set of observables of (b) with a corresponding set of observables of (a). 
\item \label{p} Two (b) models are joined and a $n$-level HEA is built, where couplings at the $n$-th level are drawn from the distribution of $n$-th level couplings of (a). 
\end{enumerate}
Angelini et al. write that ``the relation that links the exponent $\nu$ to the eigenvalue with the largest norm $\lambda_{max}$ of the linearized matrix $M$, $2^{1/\nu} = \lambda_{max}$, cannot be applied to the matrix $M$ [...] as done in [\hspace{-.2mm}\cite{castellana2019real}]. This is because in this case the eigenvalue with the largest norm is always $\lambda_{max} = \lambda_1 = 1$, the presence of this eigenvalue $\lambda_1$ is due to step [\ref{p}] of the ERG method. 
  [...] [S]tep [\ref{p}] is just introduced to iterate the procedure, given the finite size of the analyzed systems, and it is thus the step responsible for finite-size effects in the ERG results. The prediction $2^{1/\nu} = 1$ in [\hspace{-.2mm}\cite{castellana2019real}] is thus completely dominated by finite-size effects''.
In response to this statement, we observe the following: the fact that $\lambda_{\rm max} = \lambda_1 = 1$ is a feature specific to the ERG transformation, which is not present in other real-space RG approaches for the hierarchical model \cite{castellana2011real}. 
Along with its relation to finite-size effects, this feature does not give any grounds for questioning the validity of the relation $2^{1/\nu} = \lambda_{\rm max}$ specifically for the ERG transformation, i.e., on an arbitrary, case-by-case basis: In fact, the relation above can be formally derived, and is demonstrated to be valid no matter what the value of $\lambda_{\rm max}$ \cite{wilson1974the}.

In addition, we recall that the flow which follows from the ERG method \cite{angelini2013ensemble} is characterized by an early, small-$t$ regime $t \lesssim t^\ast$ (A) where $\Delta_{t,1}$ grows, and by an asymptotic, large-$t$ regime $t \gtrsim t^\ast$ (B) where $\Delta_{t,1}$ levels off, and reaches a plateau---see Fig. 6 of \cite{angelini2013ensemble} and Fig. 5 of \cite{castellana2019real}. 
Angelini et al. claim that using Eq. (\ref{eq1}) in regime (B) corresponds to considering the RG flow in an unphysical region dominated by finite-size effects, and that $\nu$ should be extracted from regime (A) \cite{angelini2019comment}. 
In this regard, we observe the following.  Instead of describing a long-wavelength regime dominated by relevant eigenvalues only, which is characteristic of systems at their critical point \cite{wilson1974the,wilson1975renormalization,justin1996quantum}, the ERG flow in (A) involves irrelevant eigenvalues, and is characterized by finite length scales \cite{justin1996quantum}---including the system size. It follows that the value of $\nu$  obtained in \cite{angelini2013ensemble} from (A) cannot describe the  exponent $\nu$, which is characteristic of a long-wavelength, critical regime where the correlation length diverges. \\
As the ERG transformation is further iterated and flows away from (A) to (B), the finite length scales of (A) are integrated out, and the RG iteration enters the long-wavelength, critical regime (B), which has been used in \cite{castellana2019real} to correctly obtain $\nu$. Importantly, we observe that the fact that (B) involves finite-size effects is a feature specific to the ERG, and gives no rationale for extracting $\nu$ from regime (A),  which is unrelated to the critical one, nor to apply Eq. (\ref{eq4}) to such regime.

\section{Large-$n$ behavior}\label{s2}

Angelini et al. claim that the plateau in (B) disappears for large $n$, see ``[i]n  [\hspace{-.1mm}\cite{angelini2013ensemble}] it was also shown that $t^\ast$ grows if larger systems are used [...],  going to $t^\ast = \infty$ in the $n \rightarrow \infty$ limit'' \cite{angelini2019comment}. However, the fact that the number of RG iterations $t^\ast$ needed to transition from (A) to (B) is an increasing function of $n$ is simply due to the fact that, the larger $n$, the more the largest norm of irrelevant eigenvalues gets close to one---see for example  Figs. 3 and 4 in \cite{castellana2019real}. In fact, the larger $n$, the larger the number of iterations $t$ needed for the irrelevant eigenvalues $(\lambda_l)^t$ to go to zero in Eq. (\ref{eq2}). 
No matter how large $n$ and $t^\ast$, regime (A) is characterized by irrelevant eigenvalues and finite characteristic lengths,  thus it does not correspond to an asymptotic critical, long-wavelength region such as (B)---see the discussion in Section \ref{s1}. As a result, the critical exponent $\nu$ determined from (A) does not represent the critical exponent related to the divergence of the correlation length, no matter how large $n$. 

As far as the large-$n$ limit is concerned, Angelini et al. also claim a correspondence between extracting $\nu$ from regime (A), and the large-$n$ limit of the ERG transformation, see ``fitting data with eq. [(\ref{eq1})] for $t < t^\ast$ [...] takes contribution not only from $\lambda_1$, but also from $\lambda_i$ with $i = 2, 3, 4$ for $n = 4$. This is exactly the correct thing to do, because, in the large $n$ limit, where step [\ref{p}] of ERG is no more needed, the matrix $M$ will not have the $\lambda_1 = 1$ eigenvalue, but only the others[, irrelevant] eigenvalues $\lambda_i$ with $2 \leq i \leq n$'' \cite{angelini2019comment}.\\ However, we observe that the argument above is not valid, for two reasons. First, while the argument claims that the large-$n$ limit is characterized by irrelevant eigenvalues only, regime (A)  is characterized by both marginally relevant and irrelevant eigenvalues: This substantial difference invalidates the claimed correspondence between extracting $\nu$ from (A) and the large-$n$ limit. 
Second, in the absence of step \ref{p}, see Section \ref{s1}, the large-$n$ limit is not characterized by irrelevant eigenvalues as claimed by Angelini et al. In fact, as depicted in Fig. 1 of \cite{angelini2013ensemble}, step \ref{p} is required in order to produce an ensemble of rescaled systems which have the same size as the original one \cite{angelini2013ensemble,angelini2019comment}: this is a fundamental property, which is closely related to the self-similarity of the RG transformation at the critical point. As a result, in the absence of step \ref{p}, and no matter what the value of $n$, the ERG transformation would map  $n$ standard deviations  $\sigma_1, \cdots, \sigma_n$ of the coupling distributions of model (a) onto $n-1$ standard deviations $\sigma'_1, \cdots, \sigma'_{n-1}$ of (b)---see steps \ref{p1}, \ref{p2} and \ref{p3} in  Section \ref{s1}. It follows that the matrix $M$ which linearizes the ERG transformation would be non-square, and the concept itself of eigenvalues $\lambda_2,  \cdots, \lambda_n$  in Angelini et al.'s argument above would be ill-defined. 

\section{Exponential fit and value of $\nu$}\label{s4}

\begin{figure}
\begin{center}
\includegraphics[scale=.9]{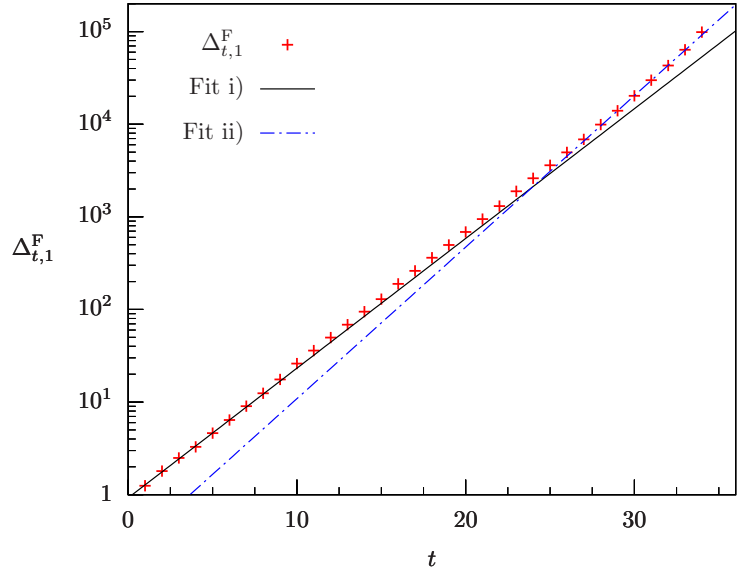}
\caption{
\label{fig1}
Difference between the ensemble renormalization group (ERG) flows of couplings at two different temperatures, $\Delta^{\rm F}_{t, 1}$, for the ferromagnetic version of the hierarchical Edwards-Anderson model (red dots) from \cite{angelini2013ensemble}, in semi-logarithmic scale. The lines represent the fit of $\log \Delta^{\rm F}_{t, 1} $ vs. $t$ with a straight line, where fit i) is made for $t \leq 6$ (black solid line), and fit ii) for $t \geq 30$ (blue dashed line). 
}
\end{center}
\end{figure}

In response to Angelini et al.'s statement ``we showed how well an exponential fit of the form [(\ref{eq1})] fits the data obtained for the ferromagnetic model with $n = 13$'', in Fig. \ref{fig1} we show the data for the ERG applied to the ferromagnetic version of the HEA, from Fig. 3 of \cite{angelini2013ensemble}. The quantity plotted in red dots, $\Delta^{\rm F}_{t, 1}$, is analogous to $\Delta_{t, 1}$ for the HEA, where `F' stands for `ferromagnetic' and the standard deviations are replaced by the values of the ferromagnetic couplings, and this quantity is plotted as a function of $t$ in an interval which belongs to regime (A). In addition, the solid and dashed lines i) and ii) have been obtained by fitting $\log \Delta^{\rm F}_{t, 1}$ vs. $t$ with a linear function for $t \leq 6$ and $t \geq 30$, respectively.
Shown in semi-logarithmic scale, the straight lines i) and ii) have different slopes, thus demonstrating that the points $\log \Delta^{\rm F}_{t, 1}$ vs. $t$ have a slope which depends on $t$. This analysis shows that, in regime (A), Eq. (\ref{eq3}) does not hold, i.e., the ERG data does not have the exponential form $\Delta^{\rm F}_{t, 1} = 2^{t/\nu}$, thus confirming the analysis at the beginning of this reply.
Given that the slope is not constant and that the fitting region used in \cite{angelini2013ensemble} to determine $\nu$ is arbitrary, the predicted value of $\nu$ from \cite{angelini2013ensemble,angelini2019comment} is also affected by arbitrariness.

Also, Angelini et al. write that the fit above has been done ``in a much broader regime $0 < t \leq 35 < t^\ast$'' compared to the corresponding fit for the HEA. In this regard we observe that, despite the fact that the interval $0 < t \leq 35 < t^\ast$ may appear to be broad, such interval still belongs to the pre-asymptotic, non-critical regime (A)---see the discussion in Section \ref{s1}.

Finally, in response to Angelini et al.'s statement that the ERG method for the ferromagnetic model ``giv[es] a value of $\nu$ that is in perfect agreement with the exact result'' \cite{angelini2019comment}, we observe that there is a $\sim 7 \%$  discrepancy between the ERG prediction $\nu = 2.076(6)$ and the value $\nu = 1.9487...$ from \cite{godina1999high}.

\vspace{1.5cm}

As a result of the analysis provided in this reply, we stand by our assertion that the ERG method yields a prediction for $\nu$ given by $2^{1/\nu}=1$ \cite{castellana2019real}. 

\ack

We thank A. Barra for valuable discussions. 

\vspace{1cm}
\bibliographystyle{unsrt}

\end{document}